\documentclass[12pt,preprint]{aastex}

\usepackage{graphics}
\usepackage{graphicx}
\usepackage{epstopdf}
\usepackage{epsfig}

\begin{document}

 \title{Long-term Photometric Analysis of the Active W UMa-type System TU Bootis}
\author{Jeffrey L. Coughlin\altaffilmark{1}}
\altaffiltext{1}{Some research funded by and performed while at Emory University}
\affil{Department of Astronomy, New Mexico State University\\P.O. Box 30001, MSC 4500, Las Cruces, New Mexico 88003-8001}
\email{jlcough@nmsu.edu}
\author{Horace A. Dale III}
\affil{Department of Physics, Emory University, Atlanta, GA 30322}
\and
\author{Richard M. Williamon}
\affil{Department of Physics, Emory University, Atlanta, GA 30322}
       
\begin{abstract}

We present multi-color light curves for the W UMa-type eclipsing binary TU Boo for two epochs separated by 22 years. An analysis of the O-C diagram indicates the earlier observations took place right in the middle of a major period change, thus allowing for a unique study on mass transfer and period changes in this W UMa-type system. We compute model fits to our light curves, along with the only other published set, using the Wilson-Devinney program, and find temporally correlated changes in the size of the secondary component with anomalies in the O-C diagram. We investigate the cause of these changes and find support for the existence of rapid, large-scale mass transfer between the components. We postulate that this interaction allows them to maintain nearly equal surface temperatures despite having achieved only marginal contact. We also find support for the evolutionary scenario in which TU Boo has undergone a mass ratio reversal in the past due to large-scale mass transfer so that what is presently the secondary component of TU Boo is in an advanced evolutionary state, oversized due to a helium-enriched core, with a total system age of $\geq$ 10 Gyr.

\end{abstract}


\section{Introduction}

The eclipsing binary TU Boo, [14$^{h}$ 04$^{m}$ 59$^{s}$, +30$\degr$ 00$\arcmin$ 00$\arcsec$, m$_{V}$ $\approx$ 12], is a short period W UMa-type system, (P $\approx$ 0.324 days), initially discovered by \citet{Guthnick26} and found to have a spectral type of G3 by \citet{Schwassmann47}. It has  only been subjected to a single published analysis in the past 50 years by \citet{Niarchos96}, who modeled the system in 1996 based on B and V light curves obtained photoelectrically in 1982. They obtained a slightly over-contact solution for the system with q = 0.498, and noted period changes and other interesting aspects of the system.

The classical theory of W UMa systems comes from the seminal work of \citet{Lucy68} in which he proposed they were zero-age contact binaries with a common convective envelope. Although the thermodynamic arguments are sound, the formation scenario encounters many observational problems. W UMa systems are virtually non-existant in young clusters, while quite populous in open clusters with ages exceeding 4-5 Gyrs, as well as globular clusters \citep{Kaluzny93,Rucinski98,Rucinski00}. As well, numerical simulations of binary formation from protostellar clouds favor early fragmentation and formation of detached systems, and seem to preclude the possibility of fission from a rapidly rotating protostar \citep{Boss93,Bonnell01}. In contrast to Lucy, a new general theory has recently arisen, \citep{Stepien06,Eker07}, which postulates that W UMa-type systems start out as detached systems with P $\approx$ 2 days and significantly different component masses. They lose angular momentum via magnetized stellar winds over several Gyr, during which time the initially more massive component evolves to terminal age main-sequence (TAMS). At this time the more massive component starts to rapidly transfer mass to the less massive, and a contact system is formed. Rapid mass transfer continues past the point of mass ratio reversal, so that what was the more massive component becomes the less massive, and visa versa. This continues until a mass ratio of $\sim$0.5 is reached, at which point a tentative equilibrium is reached and we find a typical W UMa-type system in marginal contact. Over the next several Gyr, the system undergoes thermal oscillations, although overall conservative mass transfer from the now more evolved secondary, oversized due to helium enrichment, to the primary occurs. Eventually the now more massive primary evolves to TAMS, and the stars coalesce into a single, rapidly rotating star.

\section{Observations}

For the earlier epoch, Johnson B and V observations were taken with the 36-in. Cassegrain reflector at the Fernbank Science Center Observatory in Atlanta, GA during the 1983 and 1984 observing seasons. An unrefrigerated EMI 6256s photomultiplier was used combined with a Honeywell strip chart recorder read with 5-sec timing accuracy. All observations were made differentially with respect to a comparison star and corrected for atmospheric extinction by means of nightly extinction coefficients determined from the comparison star. For the later epoch, observations in the Johnson U, B, V, R, \& I filters were taken with Lowell Observatory's 42" Hall Telescope and FLI SITe 2048$\times$2048 CCD camera, cooled by liquid nitrogen to -133$\degr$C, over five nights from April 20th-24th, 2006. Ensemble photometry was performed with respect to GSC 2012-878, 2545-811, 2545-1000, 2012-479, and 2012-831. The entirety of the photometric data for both the 1983-1984 and 2006 observations are listed in Table 1.

\section{Minimum Timings and O-C Diagram}

All observed times of minimum for our data were determined via the method of \citet{Kwee56}, and are shown in Table 2 with errors, employed filter, and type (primary or secondary eclipse). All previously published times of minima available were compiled and in cases where no error was given, which are mostly visual observations, a value of $\pm$ 0.01 days was assumed. A linear, error-weighted, least-squares fit was then performed for data after JD 2450800 and a new ephemeris calculated to be \begin{center} T$_{pri}$(HJD) = 2424609.539(5) + 0.32428316(6)$\cdot$E\end{center} where the parentheses indicate the amount of uncertainty in the last digit, and E is the epoch. An O-C diagram created using the new ephemeris is shown in Figure 1. The major period change just after JD 2446000 was noted by \citet{Niarchos96}, who calculated it to be a period decrease of 0.413 seconds, and noted that it occurred just after their 1982 observations. Figure 1 shows evidence for minor but continuous period changes afterwards, with the most noticeable change around JD 2451000. Performing weighted, least squares fits to the data before and after each major change yields a period decrease of 0.446 $\pm$ 0.012 seconds and an increase of 0.146 $\pm$ 0.013 seconds at HJD 2446085.5 and 2451216.4 respectively. It is interesting to note that while our 1983 times of minima fall well within the observed trend for other minima such as \citet{Hoffmann83}, our 1984 observations lie $\sim$0.02 days above the trend. Taking a weighted mean of the differences of our O-C values from the trend yields a value of 0.0257 $\pm$ 0.0007 days. They also happen to occur a mere 200 days before the measured date of the major period shift, well within any reasonable time window for transitionary occurrences in the system. As the only other times of minima around this epoch are a few visual observations without determined errors, whereas ours are photoelectric observations of five separate minima with well-determined errors, we take our times of minima to be real, accurate, and indicative of unusual system activity just before or at the moment of a major period change.

The sub-figure in Figure 1 is a close-up of the error-weighted averages over all filters for each recent 2006 minimum timing presented in this paper. There is a measurable difference on average of 0.00065 days, or 0.0020 phase, between the primary and secondary eclipses. As any eccentricity is extremely unlikely in a contact system, this offset is most likely due to a large star spot in the system, to be further discussed in \S4.

Inspecting the O-C residuals for times after the second major period change, there appears to be a slight upward trend indicating a constant period increase. Performing an error-weighted, least-squares quadratic fit to the O-C residuals, shown in Figure 2, yields a period increase of dP/dt = 3.46 $\pm$ 0.83 $\times$ $10^{-7}$ days yr$^{-1}$, (dP/dE = 3.07 $\pm$ 0.74 $\times$ 10$^{-10}$ days cycle$^{-1}$), which are typical values for W UMa systems \citep{Qian01,Yang03}. The implications of this trend and the other period shifts will be discussed in \S5.

\section{Light Curves and Modeling}

For our modeling, we use the 2007-08-15 version of the Wilson-Devinney code \citep{Wilson71} as implemented in the PHOEBE-0.29d package \citep{Prsa05}. Since the accuracy of the G3 spectral type determined by \citet{Schwassmann47} is unknown, we decided to use the 2MASS J, H, \& K-band magnitude measurements in order to obtain an accurate measure of the average temperature of the system. For TU Boo, these are given as J = 10.306 $\pm$ 0.023, H = 10.002 $\pm$ 0.024, and K = 9.934 $\pm$ 0.018, taken simultaneously at phase 0.35, when both components are visible. We computed color indices for J-K, J-H, and H-K, and interpolated corresponding temperatures and errors from the standard tables of \citet{Houdashelt00} and \citet{Tokunaga00}. We then performed an error-weighted mean and obtained a value of 5900 $\pm$ 150 K, matching a G0 spectral type. Thus in all our solutions we set the hotter component to T = 5900K. We set the bolometric albedos, A$_{1}$ = A$_{2}$ = 0.5, and  and the gravity darkening exponents, g$_{1}$ = g$_{2}$ = 0.32, as is customary for stars with convective envolopes \citep{Lucy67,Rucinski73}. Further support for the choice of gravity darkening exponents comes from \citet{Pantazis98}, who used a method based on the fourier analysis of light curves to observationally determine g$_{1}$ = g$_{2}$ = 0.32 $\pm$ 0.02 for the TU Boo system. For limb darkening, we used the square root law with coefficients interpolated from the tables of \citet{VanHamme93}.

As our recent 2006 CCD observations are of superior quality and number, (208 points in U and $\sim$240 in BVRI), we decided to model them before the 1983-84 observations. Inspection of the phased light curve reveals an O'Connell effect that scales with color; the magnitude difference between phase 0.25, which is brighter, and phase 0.75 ranges from 0.02$^{m}$ in I to 0.05$^{m}$ in U. This is typically explained by the presence of a spot in the system, and found by \citet{Rucinski97} to be common for contact binaries, as well as possibly related to the direction of mass transfer between components. In order to accurately obtain the orbital parameters, we first binned the data into 100 points, each separated by 0.01 phase. The binned curve was then reflected about phase 0.5 and corresponding points averaged, thus removing any asymmetries and effectively "de-spotting" the system.

As longer wavelengths show less distortion due to spots, we started our 2006 solution using only the binned and averaged R \& I light curves. Since \citet{Niarchos96} found that the components had slightly different temperatures, we used the over-contact binary not in thermal contact mode of WD, and set T$_{1}$ = 5900K. Since radial velocity curves are not available, we set the scale of the system so that the mass of the primary in solar masses equals the radius in solar radii, which leaves the secondary oversized. We consider this more physically plausible, as the other option is to leave the primary significantly undersized. Assuming a circular orbit, we let vary the secondary temperature, T$_{2}$, the mass ratio, q ($\frac{M_{2}}{M_{1}}$), the inclination, i, and primary surface potential, $\Omega$$_{1}$ = $\Omega$$_{2}$, and iterated until a satisfactory fit was found where any further corrections were less than the errors. Despite a lack of radial velocities, the mass ratio is well-constrained photometrically due to the total secondary eclipse, as supported by \citet{Mochnacki71}. In order to then obtain a spot solution, we fixed these orbital parameters and solved for a single spot on the primary using the phased, unbinned U \& B light curves. After trying many different spot configurations, we found that a hot spot near the connecting neck on the primary best fit the U \& B light curves. We then carried forth by iteration, allowing the spot latitude (Lat, ranging from 0$\degr$ at the north pole to 180$\degr$ at the south pole), the longitude (Long, ranging from 0$\degr$ to 360$\degr$, with 0$\degr$ at the inner lagrangian point, 180$\degr$ at the back end, and increasing in the direction of rotation), the angular radius (Rad, where 90$\degr$ covers exactly half the star), and the temperature factor (TF, the ratio of the spot temperature to the underlying surface temperature), as well as T$_{2}$ to vary, until a satisfactory fit was found where any further corrections were less than the errors. At this point, having both solid orbital and spot parameters, we further solved for a final solution using the U, B, V, R, \& I phased, unbinned light curves, allowing all previously mentioned parameters to vary, once again until where any further corrections were less than the errors. Our final solution nudged the hot spot slightly away from the connecting neck. In order for completeness, we then also tried solutions that employed a single hot spot on the secondary, a single dark spot on the primary, and a single dark spot on the secondary component, iterating until best-fit solutions were found each time. The solutions employing a dark spot were found to be significantly worse fits than those using a hot spot. Placing the hot spot on the primary was found to be slightly better than on the secondary, and thus we take that as our final solution. Table 3 lists our final 2006 solution, where we note errors are formal Gaussian errors outputted by the PHOEBE program, Figure 3 is a graph of our light curves with the model fit, and Figure 4 is a three-dimensional representation showing the spot location. Also, referring back to \S3, the model shows a displacement of 0.0024 phase between the primary and secondary eclipse due to the large spot, well in agreement with the observed difference of 0.0020 phase from the O-C values.

Since it is in the time period of the 1984 observations during which the O-C anomaly occurs, and as the 1983 observations are few in number, we decided to only model the 1984 observations. We found that the 1984 observations would not phase together correctly at any given value for the period, and thus we had to solve for a first time derivative of the period, which we found to be dP/dt = 9.7 $\times$ 10$^{-8}$. This is expected as the period was changing rapidly during this time period, and including the first derivative term caused the data to phase correctly. At first we tried the solution for the 2006 data, but it was found to be an inadequate fit, with or without the spot, as shown in Figure 5. The 1984 light curves do not show a noticeable O'Connell effect, and overall the relative depth of the eclipses to the shoulders is higher than the 2006 data. Moving forward with an unspotted solution, we let T$_{2}$, q, and $\Omega$$_{1}$ = $\Omega$$_{2}$ vary and iterated to a best-fit solution that ultimately had a significantly larger size of the secondary component. This is illustrated in Figure 4, Table 3 lists our values for the 1984 solution, and Figure 5 is a graph of the 1984 light curves with the model fits.

Given the numerous improvements to the WD code in the past decade, most notably the incorporation of model atmospheres, we decided to re-model the 1982 observations collected by \citet{Niarchos96}. These light curves are similar to the 2006 light curves in that they do show an O'Connell effect. In order to obtain results comparable to the 2006 solution, we allowed for one hot spot on the primary, and in addition to the spot parameters let T$_{2}$, q, and $\Omega$$_{1}$ = $\Omega$$_{2}$ vary. We found a solution with many values intermediate to those found for the 2006 and 1984 observations, but closer to the 2006 solution. The values are listed in Table 3, and the light curves with model fits are shown in Figure 6.

Although a directly measured parallax does not exist for TU Boo, and thus we have no way of directly determining its absolute magnitude, we can estimate it by employing period-color-luminosity relations (PCLR). \citet{Rucinski97b} established one such PCLR using the B-V color index. Using their relation and a value of B-V $\approx$ 0.6 for a G0V star we calculate M$_{V}$ =  4.10 $\pm$ 0.22, where the error is the mean error of the technique given by \citet{Rucinski97b}. More recently \citet{Gettel06} used the ROTSE-I data \citep{Akerlof00} to establish a much tighter relation utilizing the J-H color index, and found a value of M$_{V}$ = 4.196 $\pm$ 0.032 for TU Boo, (and for note a subsequent distance of 290 $\pm$ 4 parsecs.) These estimates of the absolute magnitude of TU Boo allow for a critical check on our models. Our model for the 2006 data yields an absolute bolometric magnitude of the system of M$_{bol}$ = 4.18. Utilizing a bolometric correction of -0.10, appropriate for a G0 star \citep{Kaler97}, gives M$_{V}$ = 4.28, which agrees well with the PCLR estimates. This validates our choice of scale, as if we were to set the scale so that the secondary star's mass and radius were equal, (in units of solar mass and radius), we find that the model would give M$_V$ = 3.86, which would be in greater disagreement with the PCLR estimates.

\section{Period Changes and Mass Transfer}

As the two components of TU Boo are in a state of marginal contact, we take the most likely cause of the period changes as the transfer of mass between the components. Referring to \citet{Kruszewski66, Plavec68, Rucinski74}, and others, we can relate the amount of mass transfer, dm, to the change in period, dP, for a system of total mass, M, via \begin{center}$\frac{dm}{dt}$ = $\frac{Mq}{3p(1-q^{2})}$$\frac{dP}{dt}$ (1) \end{center} where positive values indicate transfer from the less massive component to the more massive. Using an average value for the mass ratio of q = 0.5, and the values of dP = -0.446 $\pm$ 0.012 seconds and dP = 0.146 $\pm$ 0.013 seconds at the two major period changes results in dm = -5.44 $\pm$ 0.15 $\times$ 10$^{-6}$ M$_{\sun}$ for the first period change in 1984/1985, and dm = 1.78 $\pm$ 0.16 $\times$ 10$^{-6}$ M$_{\sun}$ for the second period change in 1999. As seen from the O-C diagram, these occurred rather rapidly, and even assuming the transfer occurred over as much as a couple years, these rates are still $\sim$10 - 100 times typical mass transfer rates for W UMa systems \citep{Yang03}. In \S4 we reported that we found a first time derivative of the period of dP/dt = 9.7 $\times$ 10$^{-8}$ for the 1984 data. Via equation 1, using q = 0.545, this yields a mass transfer rate of 4.3 $\times$ 10$^{-5}$ M$_{\sun}$ yr$^{-1}$ from the secondary to the primary, which is $\sim$10$^{2}$ - 10$^{3}$ times typical rates \citep{Yang03}. In \S3 we found that our 1984 O-C values were 0.0257 $\pm$ 0.0007 days displaced from the trend, which if we take as a sudden period increase indicates a transfer of 0.0271 $\pm$ 0.0007 M$_{\sun}$ from the secondary to the primary, which is a very large value. Thus, these very high rates of mass transfer confirm that TU Boo is a very active system capable of producing such dramatic changes as the 1984 O-C spike and increase of the size of the secondary component.

Based on the observed period changes, we propose the following scenario. Thermal relaxation oscillation models \citep{Lucy76, Flannery76, Robertson77} predict that W UMa type systems are only in a state of marginal contact and experience thermal instabilities. The secondary could have undergone a sudden change in its mass structure as a response to a critical thermal instability, which would cause it to swell in size, as observed in the 1984 model. Another possibility is a perturbation by magnetic activity \citep{Stepien06}, which could certainly be supported by the change in the O'Connell effect between 1984 and the other epochs. Either way, this sudden increase in the radius of the secondary would have resulted in a significant mass transfer to the primary, thus resulting in the observed positive spike in O-C diagram in 1984 and the first time derivative of the period needed while phasing the 1984 data. The system in that configuration though would be inherently unstable, and would quickly restore itself via the transfer of mass from the primary back to the secondary, plus some extra due to the perturbation, resulting in the overall period decrease seen after the 1984 season. In 1999, the secondary then would have experienced another instability, only smaller, resulting in a net mass transfer to the primary and thus the observed period increase. If this scenario is correct, it may show how large but rapid mass transfers back and forth between the components in W UMa systems enable large-scale mixing and thus nearly equilibrated surface temperatures while maintaining such a marginal degree of contact.

Finally, we turn our attention to the observed current period increase measured to be 3.46 $\pm$ 0.83 $\times$ 10$^{-7}$ days yr$^{-1}$, yielding a time-scale of P/\.{P} = 9.4$^{+2.9}_{-1.8}$ $\times$ 10$^{5}$ years. Using this value with q = 0.4964 in equation 1 yields a current mass flow rate from the secondary to the primary of dm/dt = 3.60 $\pm$ 0.86 $\times$ 10$^{-7}$ M$_{\sun}$ yr$^{-1}$. This is very likely correlated with the observed hot spot on the leading side of the primary, as transferred matter from the secondary would be heated in the process and spread over the observed region on the primary that the hot spot occupies. With regards to W UMa evolution, the model of \citet{Stepien06} predicts that a typical system having achieved equilibrium after mass ratio reversal would have P = 0.32 days, a = 2.3 R$_{\sun}$, M$_{1}$ = 1.08 M$_{\sun}$, R$_{1}$ = 1.02 R$_{\sun}$,  M$_{2}$ = 0.55 M$_{\sun}$, and R$_{2}$ = 0.80 R$_{\sun}$. These parameters match remarkably well our models for the TU Boo system. This stability is achieved after 6.1 Gyr since system formation, but it is only an additional 4 Gyr later that mass flow begins from the less massive secondary to the more massive primary, increasing with time as the primary evolves toward TAMS. \citet{Stepien06} gives the average value in this time period of mass transfer from secondary to primary of 6 $\times$ 10$^{-11}$ $M_{\sun}$ yr$^{-1}$, thus our currently observed rate at 6 $\times$ 10$^{3}$ of this value indicates that the primary component of TU Boo may be very close to TAMS, and the age of the entire system is $\geq$ 10 Gyr. This state would also explain the volatile nature of the system and the enhanced mass transfer rates we observe.

\section{Discussion and Conclusion}

We have complied a comprehensive O-C diagram for the past $\sim$80 years showing major period changes in the TU Boo system, with photometric data obtained at the time of a major period change. We postulate that these changes are be caused by mass transfer between the components due to thermal oscillations resulting from marginal contact. We interpret the 1984 O-C spike, temporally correlated with the increase in the size of the secondary, as evidence for rapid mass-exchange between components in this, and thus likely other W UMa-type systems. This may very well explain how these systems can maintain near-equal surface temperatures despite such marginal degrees of contact. The well-constrained scale of the system due to physical considerations and absolute magnitude calculations validates the results for absolute masses and radii of the components. Along with the observed current mass transfer rate, this strongly supports the theory that secondary stars in W UMa systems are evolved and oversized due to helium-enriched cores.

\acknowledgments
\textbf{Acknowledgments.} The authors would like to thank the anonymous referee for comments which helped to greatly improve the paper. The authors would also like to thank Lowell Observatory for use of the 42" Hall telescope for the 2006 observations, and the Scholarly Inquiry and Research at Emory (SIRE) program for travel funding. Jim Sowell at the Georgia Institute of Technology is thanked for digitizing the 1983/84 data. Additionally great appreciation is extended to the many amateur astronomers who have collected times of minima. This publication makes use of data products from the Two Micron All Sky Survey, which is a joint project of the University of Massachusetts and the Infrared Processing and Analysis Center/California Institute of Technology, funded by the National Aeronautics and Space Administration and the National Science Foundation.

\clearpage

\begin{deluxetable}{ccc}
\tablewidth{0.0pt}
\tablenum{1}
\tablecaption{\sc{1983-1984 \& 2006 Photometric Data}}
\tablecolumns{3}
\tablehead{HJD & Filter & Differential\\&&Magnitude}
\startdata
2445485.62928 & V & 0.3346\\
2445485.63018 & B & 0.3169\\
2445485.63511 & V & 0.1376\\
2445485.63570 & B & 0.1402\\
2445485.64447 & V & 0.0295\\
2445485.64528 & B & 0.0726\\
\nodata & \nodata & \nodata\\
2453845.72513 & B & 12.2663\\
2453845.73038 & V & 12.0153\\
2453845.73575 & R & 10.9673\\
2453845.73847 & I & 11.3302\\
2453845.74106 & B & 12.3927\\
2453845.74383 & V & 12.1380\\
2453845.74627 & R & 11.0853\\
2453845.74891 & I & 11.4608\\
\nodata & \nodata & \nodata\\
\enddata
\tablecomments{Table 1 is published in its entirety in the electronic edition of the journal. A portion is shown here for guidance regarding its form and content.} 
\end{deluxetable}

\begin{deluxetable}{cccc}
\tablewidth{0.0pt}
\tablenum{2}
\tablecaption{\sc{Observed Times of Minimum}}
\tablecolumns{4}
\tablehead{T$_{min}$ (HJD) & Error $\pm$ & Filter & Type}
\startdata
2445487.719465 & 0.000511 & B & Sec\\
2445487.719982 & 0.000361 & V & Sec\\
2445791.761792 & 0.001513 & V & Pri\\
2445791.762665 & 0.000310 & B & Pri\\
2445793.708029 & 0.000540 & B & Pri\\
2445793.708405 & 0.000541 & V & Pri\\
2445798.730438 & 0.000626 & B & Sec\\
2445798.734353 & 0.000661 & V & Sec\\
2445810.731517 & 0.000266 & B & Sec\\
2445810.731730 & 0.000537 & V & Sec\\
2445817.704999 & 0.000345 & B & Pri\\
2445817.705160 & 0.000324 & V & Pri\\
2453845.774816 & 0.001091 & I & Sec\\
2453845.775159 & 0.003841 & B & Sec\\
2453845.775320 & 0.001533 & V & Sec\\
2453845.936530 & 0.003593 & V & Pri\\
2453845.936557 & 0.000638 & B & Pri\\
2453845.936603 & 0.000877 & R & Pri\\
2453845.937036 & 0.001312 & I & Pri\\
2453846.747825 & 0.000674 & I & Sec\\
2453846.747921 & 0.001766 & B & Sec\\
2453846.747934 & 0.000485 & V & Sec\\
2453846.748038 & 0.000146 & R & Sec\\
2453846.748191 & 0.000839 & U & Sec\\
2453847.720438 & 0.000608 & U & Sec\\
2453847.720559 & 0.001036 & R & Sec\\
2453847.720962 & 0.001083 & I & Sec\\
2453847.721135 & 0.000169 & B & Sec\\
2453847.721415 & 0.001085 & V & Sec\\
2453847.881942 & 0.000149 & U & Pri\\
2453847.882126 & 0.000729 & B & Pri\\
2453847.882361 & 0.000591 & V & Pri\\
2453847.882465 & 0.000760 & R & Pri\\
2453847.882470 & 0.001102 & I & Pri\\
2453848.693499 & 0.000758 & I & Sec\\
2453848.693662 & 0.000845 & B & Sec\\
2453848.693705 & 0.000596 & R & Sec\\
2453848.693718 & 0.000375 & V & Sec\\
2453848.693779 & 0.000365 & U & Sec\\
2453848.855010 & 0.000266 & U & Pri\\
2453848.855100 & 0.000917 & B & Pri\\
2453848.855258 & 0.000145 & V & Pri\\
2453848.855379 & 0.000506 & I & Pri\\
2453848.855407 & 0.000301 & R & Pri\\
2453849.666273 & 0.000357 & V & Sec\\
2453849.666320 & 0.001623 & I & Sec\\
2453849.667217 & 0.001572 & U & Sec\\
2453849.827944 & 0.000761 & B & Pri\\
2453849.828036 & 0.000175 & U & Pri\\
2453849.828067 & 0.000341 & V & Pri\\
2453849.828159 & 0.000143 & R & Pri\\
2453849.828247 & 0.000089 & I & Pri\\
\enddata
\end{deluxetable}

\begin{figure}
\centering
\epsfig{file=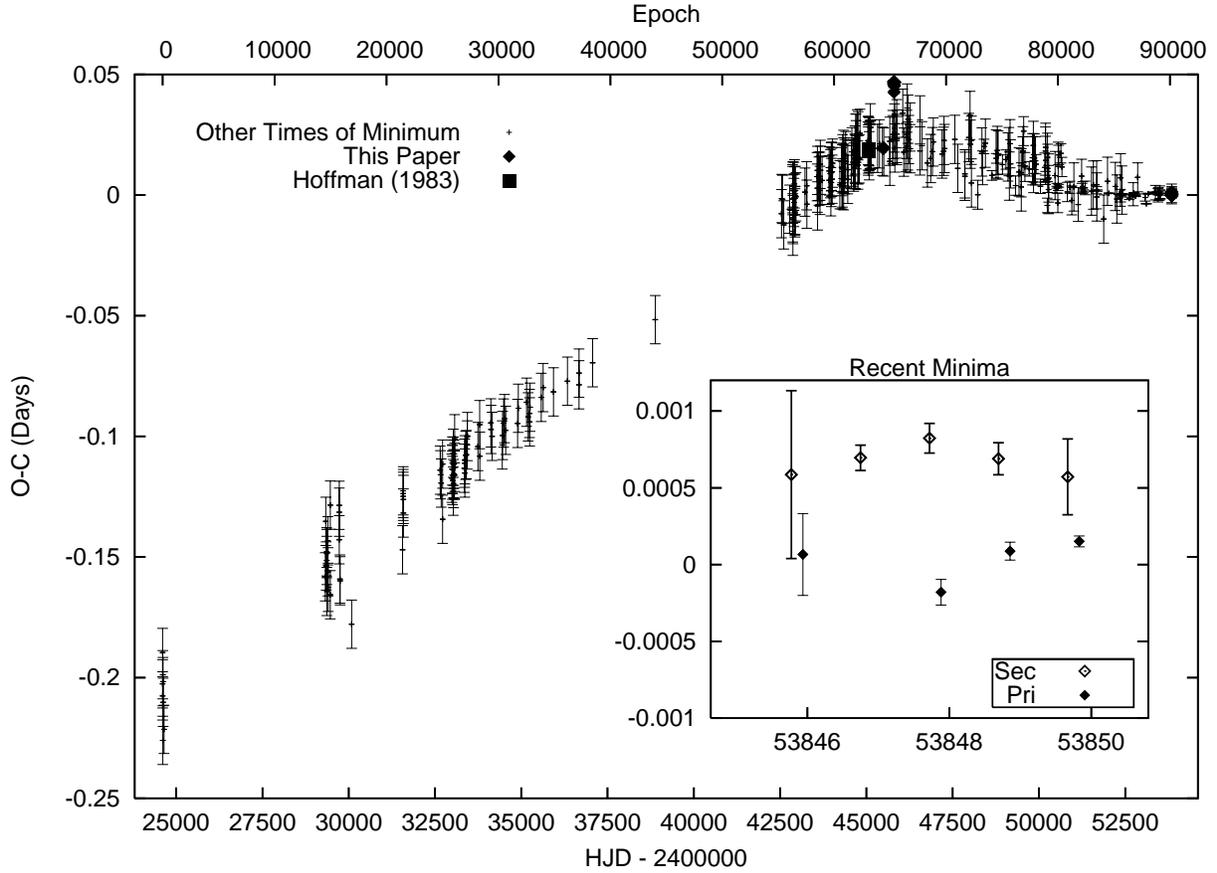, angle=0, width=\linewidth}
\caption{O-C Diagram for all compiled times of minima. Minima from this paper are shown in diamonds, and minima from \citet{Hoffmann83}, which are used in \citet{Niarchos96}, are shown in squares. Note that major period shifts occur at HJD 2446000 and 2452000. The figure in figure is a close-up of the recent times of minima from this paper. Note the slight offset between the primary and secondary minima, due to the large star spot in the system.}
\end{figure}

\begin{figure}
\centering
\epsfig{file=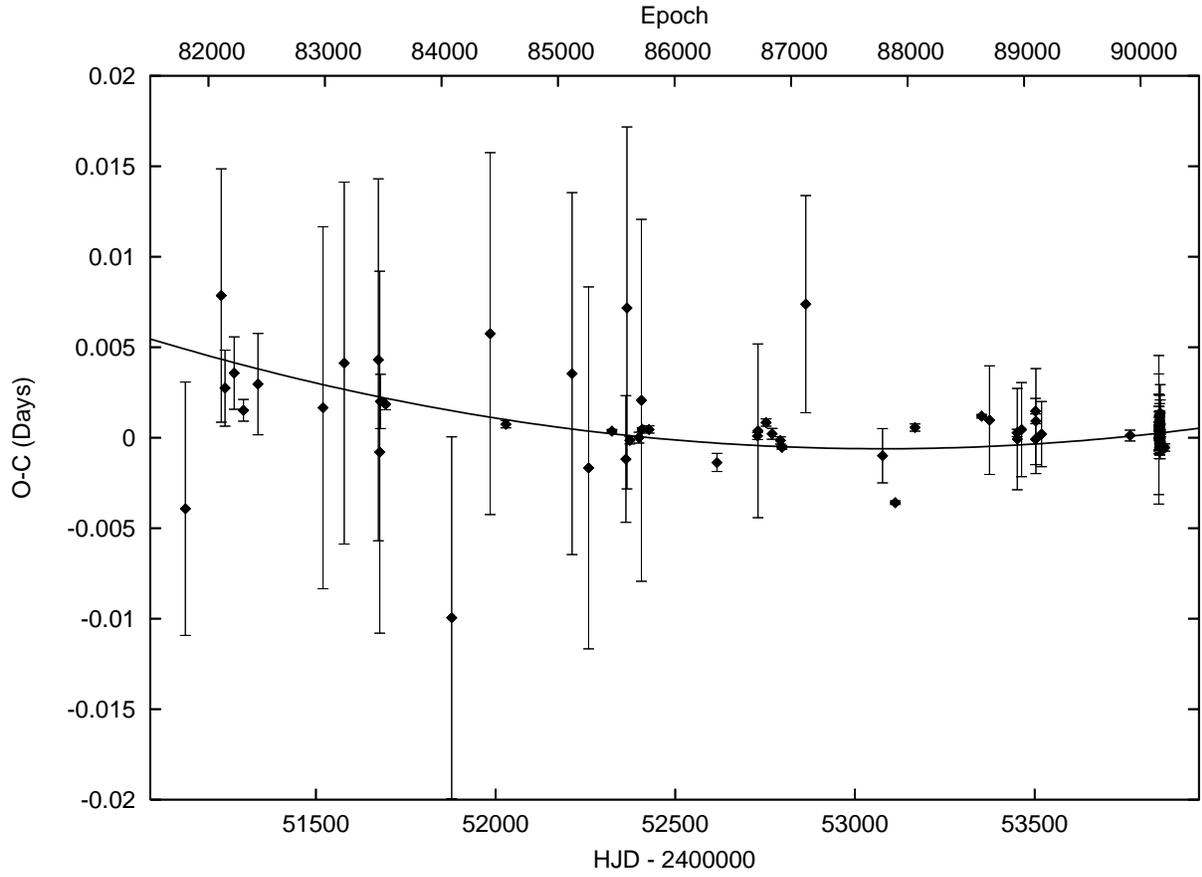, angle=0, width=\linewidth}
\caption{Quadratic fit to O-C residuals after the second major period change, indicating a current period increase of dP/dt = 3.46 $\pm$ 0.83 $\times$ $10^{-7}$ days yr$^{-1}$.}
\end{figure}

\begin{deluxetable}{lcccc}
\tablewidth{0.0pt}
\tablenum{3}
\tablecaption{\sc{Orbital Solutions for the 2006, 1984, and 1982 Light Curves.}} 
\tablecolumns{5}
\tablehead{Parameter & Symbol & 2006 Values & 1984 Values & 1982 Values}
\startdata
Inclination ($\degr$) & i & 89.2 $\pm$ 1.2 & 89.2 $\pm$ 2.0 & 89.2 $\pm$ 6.7 \\
Primary Temperature (K) & T$_{1}$ & 5900\tablenotemark{*} & 5900\tablenotemark{*} & 5900\tablenotemark{*} \\
Secondary Temperature (K) & T$_{2}$ & 5870 $\pm$ 2 & 5877 $\pm$ 7 & 5899 $\pm$ 20 \\
Mass Ratio (M$_{2}$/M$_{1}$) & q & 0.4964 $\pm$ 0.0007 & 0.545 $\pm$ 0.004 & 0.51 $\pm$ 0.01  \\
Surface Potential & $\Omega$$_{1}$ = $\Omega$$_{2}$ & 2.8365 $\pm$ 0.0019 & 2.916 $\pm$ 0.009 & 2.816 $\pm$ 0.022\\
Semi-Major Axis (R$_\sun$) & a & 2.29\tablenotemark{*} & 2.29\tablenotemark{*} & 2.29\tablenotemark{*} \\
Eccentricity & e & 0.0\tablenotemark{*} & 0.0\tablenotemark{*}  & 0.0\tablenotemark{*} \\
Fractional Radius of Primary & r$_{1}$ & 0.450 & 0.445 & 0.459\\
Fractional Radius of Secondary & r$_{2}$ & 0.328 & 0.341 & 0.345\\
Mass of Primary (M$_\sun$) & M$_{1}$ & 1.027 & 0.995 & 1.017 \\ 
Mass of Secondary (M$_\sun$) & M$_{2}$ & 0.510 & 0.542 & 0.520 \\
Radius of Primary (R$_\sun$) & R$_{1}$ & 1.030 & 1.020 & 1.050 \\
Radius of Secondary (R$_\sun$) & R$_{2}$ & 0.750 & 0.780 & 0.790 \\
Bolometric Mag of Primary & M$_{bol,pri}$ & 4.63 & 4.660 & 4.58 \\
Bolometric Mag of Secondary & M$_{bol,sec}$ & 5.34 & 5.260 & 5.22 \\ 
Albedo (bolometric) & A$_{1}$ = A$_{2}$ & 0.5\tablenotemark{*} & 0.5\tablenotemark{*}  & 0.5\tablenotemark{*} \\ 
Gravity Darkening & g$_{1}$ = g$_{2}$ & 0.32\tablenotemark{*} & 0.32\tablenotemark{*} & 0.32\tablenotemark{*} \\
Spot Temperature Factor  & TF & 1.023 $\pm$ 0.001 & - & 1.035 $\pm$ 0.023  \\
Spot Angular Radius ($\degr$)  & Rad & 37.7 $\pm$ 2.4 & - & 31.5 $\pm$ 7.6 \\
Spot Latitude ($\degr$)  & Lat & 82.7 $\pm$ 6.1 & - & 107.8 $\pm$ 17.8 \\
Spot Longitude ($\degr$)  & Long & 290.2 $\pm$ 1.3 & - & 260.3 $\pm$ 8.5 \\
\enddata
\tablenotetext{*}{Fixed}
\tablecomments{The errors listed are the formal Gaussian errors given by \sc{phoebe}.}
\end{deluxetable}

\begin{figure}
\centering
\epsfig{file=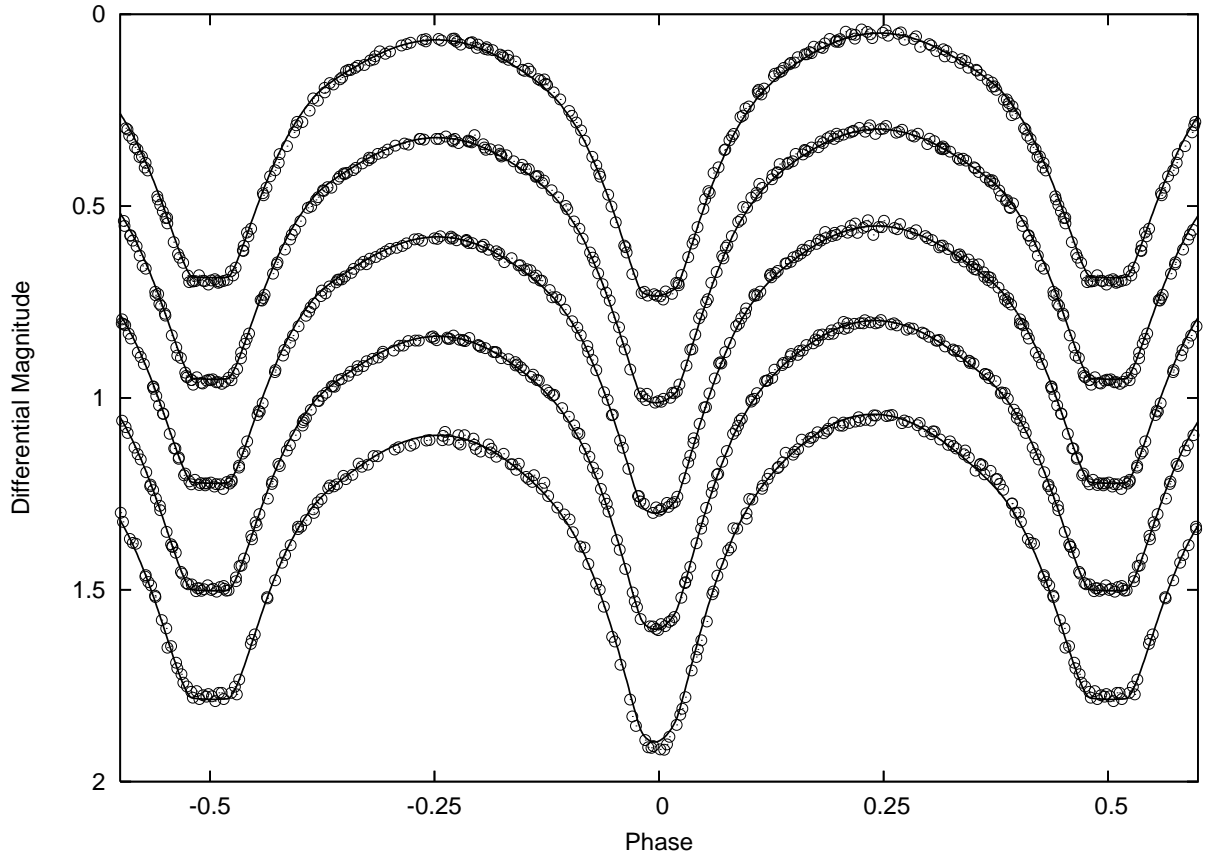, angle=0, width=\linewidth}
\caption{2006 Data with Model Fits. From Top to Bottom: I, R, V, B, U Filters.}
\end{figure}

\begin{figure}
\centering
\epsfig{file=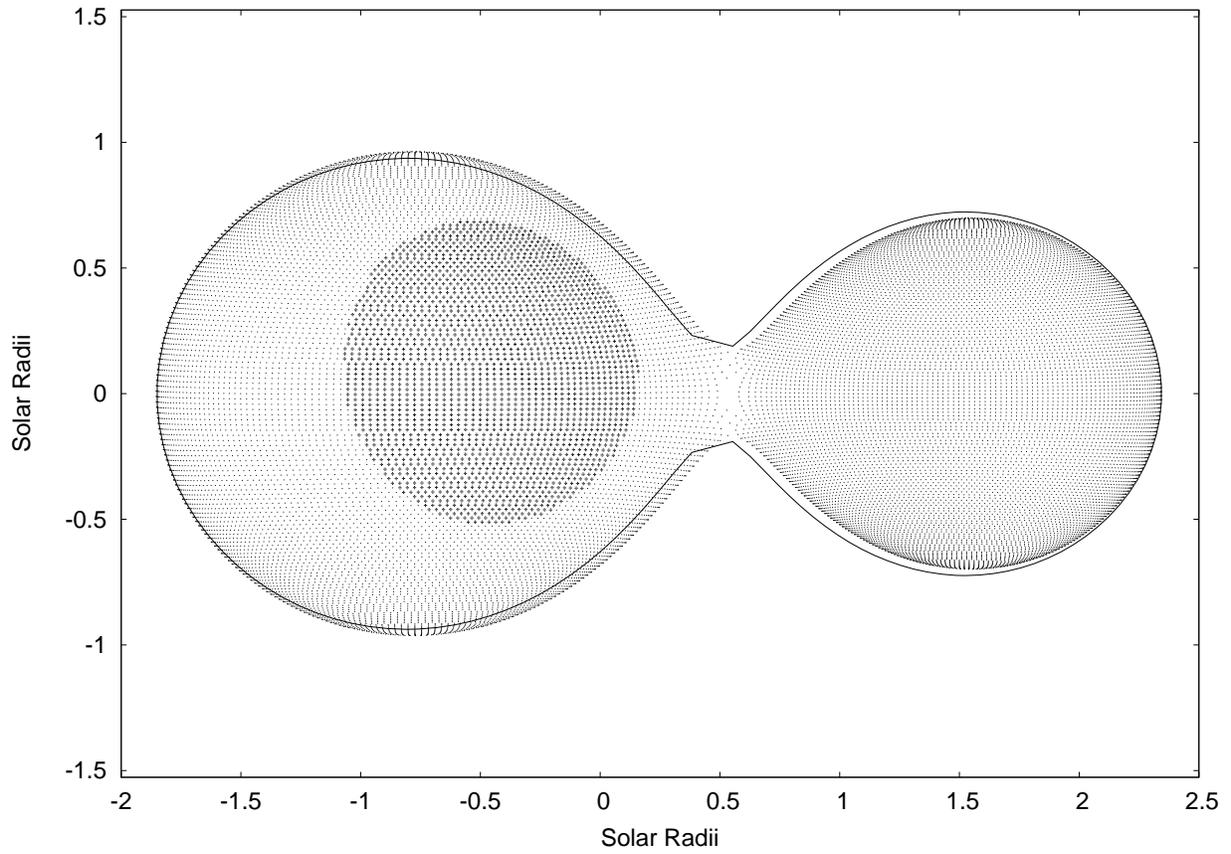, angle=0,width=\linewidth}
\caption{3D Model of TU Boo at Phase 0.25 based on the 2006 model. The spot shown is a hot spot $\sim$130K above the surface temperature. The solid line is the outline of the 1984 model, showing the increase in the size of the secondary star that occurred during that epoch.}
\end{figure}

\begin{figure}
\centering
\epsfig{file=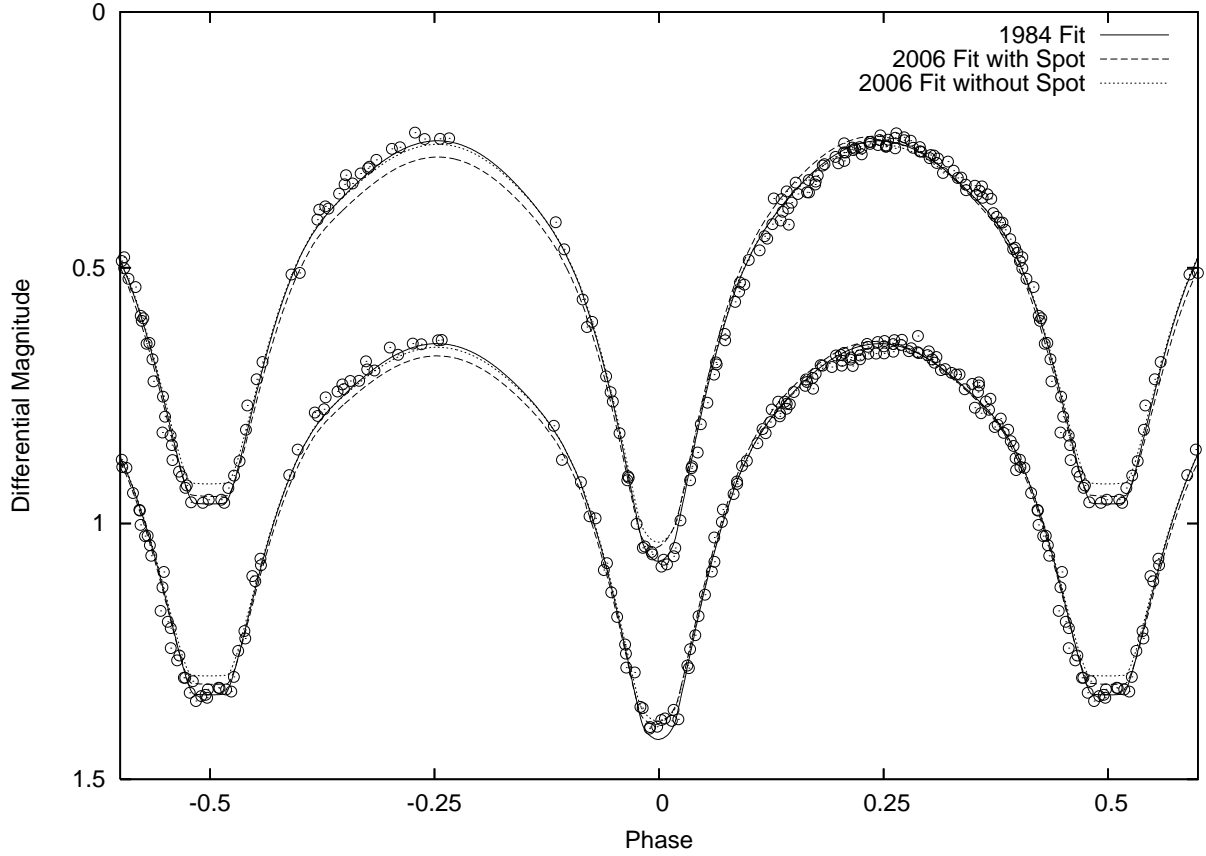, angle=0, width=\linewidth}
\caption{1984 Data with Model Fits, with the V filter on top and the B filter below. The models for the 2006 orbital solutions, with and without a spot, are shown by the dashed and dotted line respectively. Note that the 2006 spotted solution fails to reproduce the height of the shoulder at phase -0.25, and the 2006 unspotted solution fails to reproduce the depth of the secondary minima.}
\end{figure}

\begin{figure}
\centering
\epsfig{file=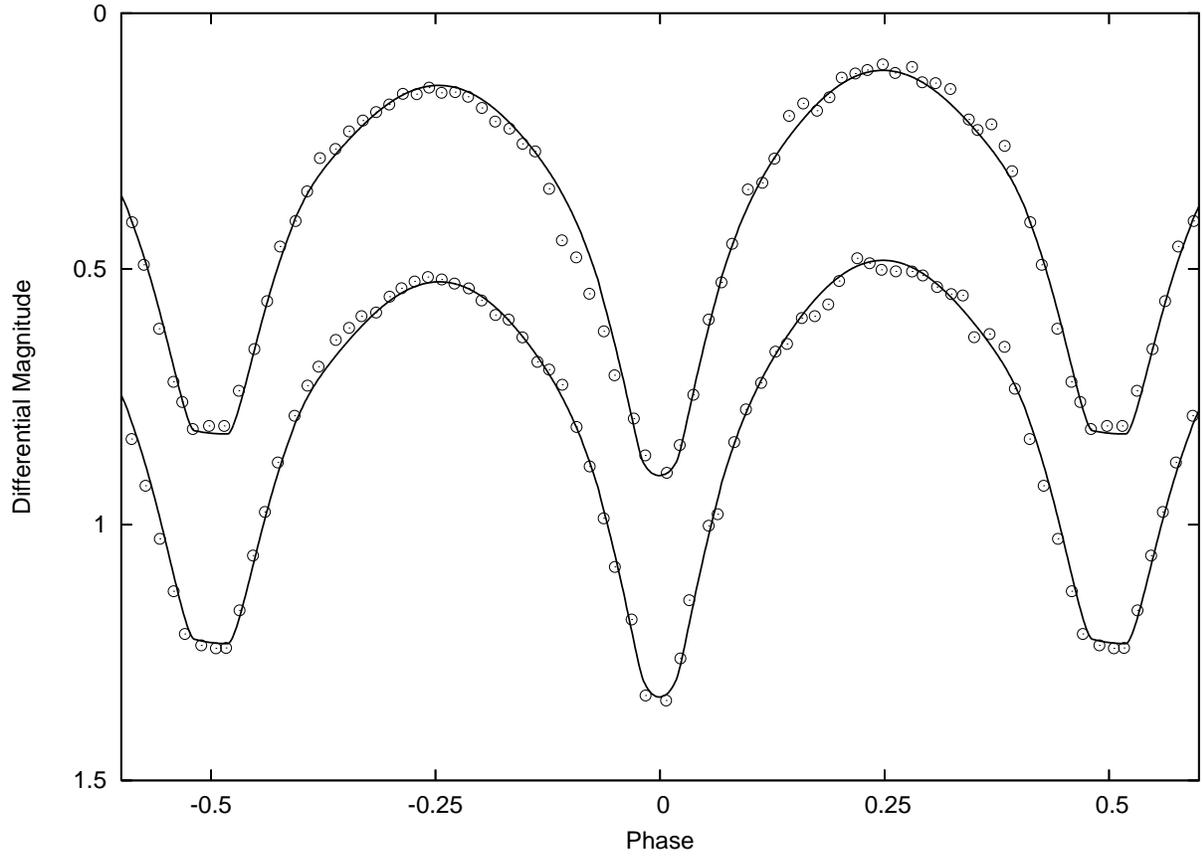, angle=0, width=\linewidth}
\caption{1982 Data from \citet{Niarchos96} with Model Fits, with the V filter on top and the B filter below.}
\end{figure}

\end{document}